\documentclass{article}

\usepackage{neurips_2023}

\usepackage[utf8]{inputenc} 
\usepackage[T1]{fontenc}    
\usepackage{hyperref}       
\usepackage{url}            
\usepackage{booktabs}       
\usepackage{amsfonts}       
\usepackage{nicefrac}       
\usepackage{microtype}      
\usepackage{xcolor}         
\usepackage{prettyref}
\newrefformat{fig}{Figure~\ref{#1}}
\newrefformat{chap}{Chapter~\ref{#1}}
\newrefformat{alg}{Algorithm~\ref{#1}}
\newrefformat{tab}{Table~\ref{#1}}
\usepackage{amsmath}
\usepackage{amsfonts}
\usepackage{amsthm}
\usepackage{amssymb}
\usepackage{mathrsfs}
\usepackage{bbm}
\usepackage{algorithm}
\usepackage{algpseudocode}
\usepackage{graphicx}
\usepackage{wrapfig,lipsum,booktabs}

\DeclareMathOperator*{\argmax}{argmax}
\DeclareMathOperator*{\argmin}{argmin}

\newcommand{\ugwass}[2]{\mathcal{GW}^\mathbf{U}(#1,#2)}

\title{EMPOT: partial alignment of density maps and rigid body fitting using unbalanced Gromov-Wasserstein divergence}

\author{%
  David S.~Hippocampus\thanks{Use footnote for providing further information
    about author (webpage, alternative address)---\emph{not} for acknowledging
    funding agencies.} \\
  Department of Computer Science\\
  Cranberry-Lemon University\\
  Pittsburgh, PA 15213 \\
  \texttt{hippo@cs.cranberry-lemon.edu} \\
}

\begin{document}

\maketitle

\begin{abstract}
Aligning EM density maps and fitting atomic models are essential steps in single particle cryogenic electron microscopy (cryo-EM), with recent methods leveraging various algorithms and machine learning tools. As aligning maps remains challenging in the presence of a map that only partially fits the other (e.g. one subunit), we here propose a new procedure, EMPOT (EM Partial alignment with Optimal Transport), for partial alignment of 3D maps. EMPOT first finds a coupling between 3D point-cloud representations, which is associated with their so-called unbalanced Gromov Wasserstein divergence, and second, uses this coupling to find an optimal rigid body transformation. Upon running and benchmarking our method with experimental maps and structures, we show that EMPOT outperforms standard methods for aligning subunits of a protein complex and fitting atomic models to a density map, suggesting potential applications of Partial Optimal Transport for improving Cryo-EM pipelines.
\end{abstract}

\section{Introduction}
\subsection{Background}

3D alignment of density maps is an important step before comparing protein structures solved from cryogenic electron microscopy (cryo-EM). In this context, various computational methods have been developed to register EM maps i.e., to find a rigid body transformation (rotation and translation) that aligns the maps \cite{han2021vesper,harpaz2023three,riahi2022alignot, singer2023alignment}. However, producing an automated and accurate alignment is still challenging when one map only matches a portion of the other, requiring algorithms that can perform partial registration from large 3D grids with different levels of intensities (typically $\sim 100^3$ to $\sim 500^3$ voxels). Similarly, developing accurate methods to partially align atomic structures within a larger map is needed to efficiently build protein complex models. This process generally involves a first step of rigid-body fitting that is similar to registering 3D maps, followed by local refinement (flexible fitting) \cite{alnabati2019advances, kawabata2018rigid}, with recent efforts leveraging machine learning \cite{demoem, embuild, deeptracer} to improve and/or automate traditionally used manual placement and exhaustive search methods.

\subsection{Main Contributions}

This paper has three main contributions. (1) We introduce a new method, called \textbf{EMPOT} (EM Partial alignment with Optimal Transport), that relies on the theory of Optimal Transport (OT) and the use of the so-called Unbalanced Gromov-Wasserstein divergence \cite{sejourne2021unbalanced}, to compare distributions with different total mass. Our method relies on using point-cloud representations of the maps, from which we compute a coupling associated with this divergence and derive an optimal rotation and translation. (2) We benchmark EMPOT with standard and recent methods of density maps alignment, showing superior performance for handling alignment of partial maps. (3) We run EMPOT to fit an atomic model structure predicted by AlphaFold \cite{jumper2021highly} to a density map, also showing superior performance 
compared with other methods designed for rigid body fitting, and suggesting some another potential application of our framework for atomic model building. 

\subsection{Related work}

In recent years, several methods have been introduced to improve standard alignment methods for cryo-EM \cite{chimeraX}, by doing alignment in Fourier space \cite{han2021vesper,harpaz2023three}, or by optimizing optimal transport-based distances \cite{riahi2022alignot, singer2023alignment}. Compared with these transport-based methods, the optimization problem that we formulate here relies on the Gromov Wasserstein divergence \cite{sejourne2021unbalanced}, which has previously been used in various other registration problems \cite{solomon2016entropic}. In particular, it does not assume that particles carry the same mass, while using intrinsic distances for each density map, instead of defining a cost function between them. 

Upon adapting our procedure to partially fit an atomic model with a density map, we can also relate EMPOT with rigid body fitting methods used for model building. In this context, our method is specifically suited for so-called single local (partial fit of a single subunit) or multiple global (with several subunits) fitting problems \cite{kawabata2018rigid}. While a wide range of descriptors have been proposed, including secondary structure elements \cite{phenix}, gaussian mixture models \cite{gmfit}, feature points \cite{zhang2010fast} and neural networks \cite{demoem, embuild, deeptracer}, our method is the first, to our knowledge, to apply a transport based metric on a large point cloud representing both maps and models. Similar to the experiment performed in this paper, EMBuild \cite{embuild} also relies on using AlphaFold \cite{jumper2021highly} for generating atomic structures, that are subsequently fitted with a density map in Fourier space. In comparison, our method avoids the need for curated datasets and extensive neural network training, with the direct use of point cloud aligned by partial optimal transport.

\section{Methods}
\subsection{Unbalanced Gromov-Wasserstein divergence}\label{sec:backgroundOT}
 We briefly provide some background on the Gromov-Wasserstein divergence \cite{sejourne2021unbalanced}.
For two given sets of 3D points $\mathbf{A} = \{a_1,\dots,a_n\}$ and $\mathbf{B} = \{b_1,\dots,b_m\}$, we consider the distributions $\alpha = \sum_{i=1}^m \alpha_i \delta_{a_i}$ and $\beta =  \sum_{j=1}^n \beta_j \delta_{b_j}$, where $\alpha_i = \frac{1}{n}, \beta_j = \frac{1}{m}$ and $\delta_x$ is a Delta Dirac function at $x$. For each distribution, we also define an inner cluster cost matrix $C^\alpha_{i,j} = d(a_i,a_j)^2$ for $\mathbf{A}$, and $C^\beta_{i,j} = d(b_i,b_j)^2$ for $\mathbf{B}$, where $d$ is the Euclidean distance. The \emph{unbalanced Gromov-Wasserstein divergence} \cite{sejourne2021unbalanced} between $\mathbf{A}$ and $\mathbf{B}$, denoted by $\ugwass{\mathbf{A}}{\mathbf{B}}$ is then given by




 \begin{equation}
\ugwass{\mathbf{A}}{\mathbf{B}} = \left[ \min_{P \in \mathbb{R}_{\geq 0}^{n \times m}} \quad  \sum_{i,j=1}^n\sum_{k,l=1}^m \lVert C^\alpha_{i, j} - C^\beta_{k,l}\rVert^2P_{i, k}P_{j,l}+ \rho \left[\text{KL}^\otimes (\pi_1|\alpha) + \text{KL}^\otimes (\pi_2|\beta) \right] \right]^{1/2},
\label{eq:ugwass-definition}
\end{equation}

where $\rho >0$ is the so-called the \emph{unbalanced parameter}, $\pi_1$ and $\pi_2$ are marginal distribution constraints $\pi_{1,j} = \sum_i P_{j,i}, \pi_{2,j} = \sum_i P_{i,j}$, and $\text{KL}^\otimes$ is the \emph{quadratic Kullback-Leibler divergence}, so that $\text{KL}^\otimes(\pi_1|\alpha) = \sum_{i,j} \log (\frac{\pi_{1,i}\pi_{1,j}}{\alpha_i\alpha_j})\pi_{1,i}\pi_{1,j} - \sum_{i,j} \pi_{1,i}\pi_{1,j} + \sum_{i,j}\alpha_i\alpha_j$. 
The matrix $P$ that is the minimizer of \eqref{eq:ugwass-definition} then defines an optimal transport plan. This formulation 
relaxes the \emph{Gromov-Wasserstein distance} \cite{memoli2011gromov} (where $P$ satisfies marginal constraints), and makes it suitable for the case where $\mathbf{A}$ and $\mathbf{B}$ do not match entirely. In practice, this divergence and the associated transport plan can be efficiently approximated using 
entropic regularization \cite{sejourne2021unbalanced}.

\subsection{Procedure for partial alignment of EM maps}
Given two 3D EM maps ($\mathcal{A}$ and $\mathcal{B}$), we now describe our procedure to align them (the algorithm is also summarized in Appendix A of our SI file). First, we represent $\mathcal{A}$ and $\mathcal{B}$ as 3D point clouds $\mathbf{A} = \{a_1,\dots,a_n\}, \mathbf{B} = \{b_1,\dots,b_m\}$ respectively, by using the topology representing network algorithm (TRN) \cite{martinetz1994topology} (similar to other recent methods that process EM density maps \cite{trn2021,riahi2022alignot}). Next, we compute the transport plan $P$  for $\mathcal{GW}^U(\mathbf{A}, \mathbf{B})$, as defined in section \ref{sec:backgroundOT} and use it to get a map $\pi$ of all the points from $\mathbf{A}$ to $\mathbf{B}$, such that
\begin{equation}
    \pi(a_i) = b_{\argmax_j P_{i,j}}.
\end{equation}
To extract a rotation and translation from $\pi$, we finally minimize the root mean squared deviation (RMSD) between paired points, as
\begin{equation}
    \label{eq:loss-ls-definition}
    R_{opt},T_{opt} = \argmin_{R,T} \sum_{i=1}^n \lVert Ra_i + T - \pi(a_i) \rVert^2.
\end{equation}
This optimization problem can be explicitly solved using the \emph{Kabsch algorithm} \cite{kabsch1976solution}, as described in details in our Supplementary Information file, Appendix B. 



\subsection{Implementation}

We implemented EMPOT in Python 3.10. To sample a point cloud representation of an EM map using TRN, we adapted code from ProDy \cite{prody2021}, using the same hyperparameters as in \cite{riahi2022alignot}. To compute the unbalanced Gromov-Wasserstein divergence, we used the code from \cite{sejourne2021unbalanced}, with $\epsilon=2000$ and $\rho=10^5$. We used the NumPy package for matrix operations of the Kabsch algorithm. The point
cloud size was set to 500 in our first experiment and 2000 in the second. Visualization of maps and structures was done using UCSF ChimeraX \cite{chimeraX}. Code and datasets will be available upon publication.

\subsection{Datasets} \label{sec:dataset}

\begin{wrapfigure}{r}{4cm}
\centering
    \includegraphics[width=.65\linewidth]{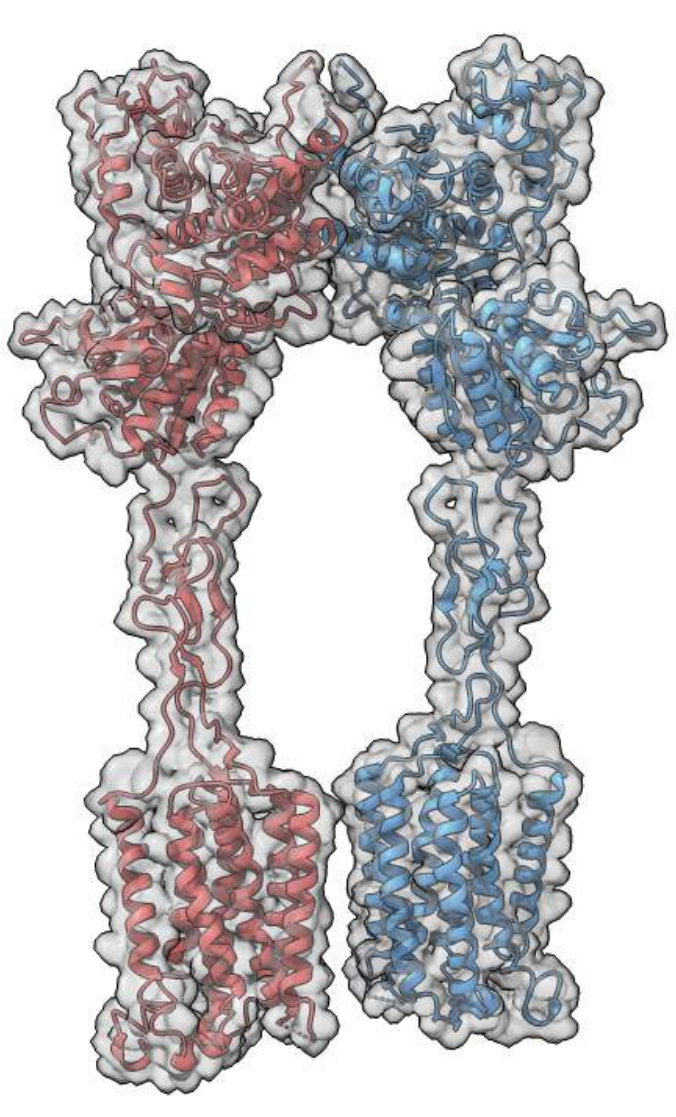}
  \caption{3D structure of the Metabotropic Glutamate Receptor 5 Apo Form complex.}
  \label{fig:maps}
\end{wrapfigure}

In our main experiments, we used an atomic cryo-EM structure of the Metabotropic Glutamate Receptor 5 Apo Form complex and its associated density map (PDB:6N52 and EMD:0346) \cite{koehl2019structural}, shown in Figure \ref{fig:maps}. 
This complex consists of two identical chains which we used for partial alignment with EMPOT, by using the \texttt{molmap} command in UCSF ChimeraX \cite{chimeraX} to generate a density map from a 3D structure. The structure from PDB:5fn5 \cite{bai2015sampling} (Figure S1) was also used and similarly processed in SI file.  


\section{Results}

\subsection{Partial alignment of density maps}

\begin{figure}
  \centering
    \includegraphics[width=.99\linewidth]{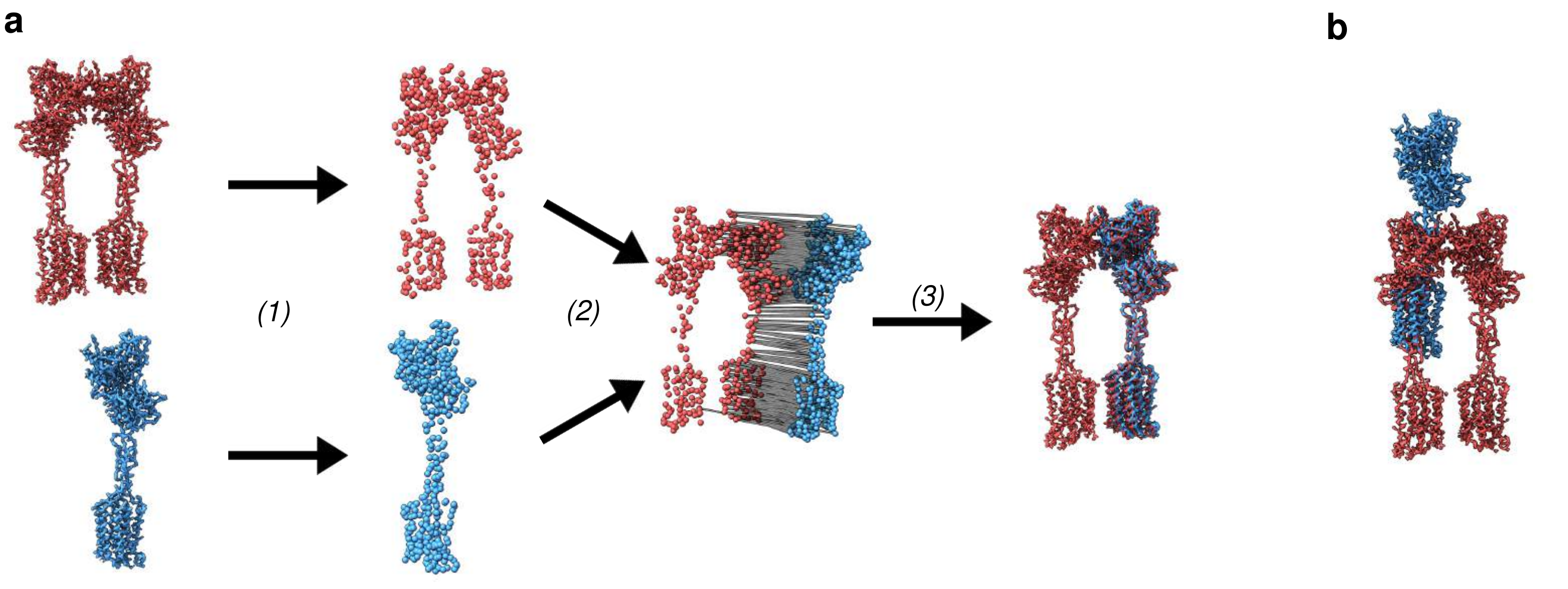}
  \caption{\textbf{(a)} Illustration of partial alignment with our method in three steps. \textit{(1)} Converting density maps to point clouds using TRN. \textit{(2)} Finding corresponding points by computing the unbalanced Gromov-Wasserstein distance transport plan. \textit{(3)} Calculating the optimal rotation and translation using the Kabsch algorithm. \textbf{(b)} Illustration of an unsuccessful alignment using the built-in \texttt{fitmap} command in UCSF ChimeraX (Results from other methods are shown in Supplementary figure S1).}
  \label{fig:exp1}
\end{figure}

To test our method, we processed the cryo-EM structure from our dataset (see section \ref{sec:dataset}). 
Upon converting the single subunit structure into a density map at 4\AA \ using the \texttt{molmap} function in ChimeraX \cite{chimeraX}, we aligned it to the golobal map using EMPOT, with the result and intermediate steps (point cloud generation and matching with the Gromov Wasserstein divergence) shown in Figure \ref{fig:exp1}a. Overall, our procedure was successful at partially aligning the map to one of the two subunits, while the \texttt{fitmap} command from UCSF ChimeraX (that minimizes their correlation) fails to do so, as shown in Figure \ref{fig:exp1}b.

We further benchmarked the performance of EMPOT against other methods that include a couple of OT-based methods, AlignOT \cite{riahi2022alignot} and BOTalign \cite{singer2023alignment}, as well as another recent method that uses Fast Fourier transform, EMAlign \cite{harpaz2023three}, and UCSF ChimeraX's \texttt{fitmap} command \cite{chimeraX}. Note that since the complex consists of two identical chains, there are two ground truth alignments. For each measurement, we thus reported the best of the two possible angle differences of rotation, as well as the RMSD's obtained from matching the output atoms with the reference. For methods that are potentially sensitive to the initial impositions, we also used different random impositions (5 for AlignOT, ChimeraX, 2 for EMPOT) and selected the overall best output. 
The results, shown in Table \ref{table:empot}, indicate a significant performance improvement from our method in both metrics. They can also be explained, as some methods (AlignOT and BOTalign) are not designed to work with partial density maps, while others (EMAlign and ChimeraX) are prone to getting stuck in local minima. Supplementary Figure S1 illustrates a more detailed chart of the results, with representative examples that illustrate these issues.
We also note that with an average runtime of $181.13$ seconds, on a 12th Gen Intel(R) Core(TM) i5-1240P 1.70 GHz CPU, our method can be performed on a standard workstation in reasonable time. Using an alternative dataset (with four heterogeneous chains) produced similar results, as detailed in SI file, Appendix C and Table S1.

\begin{table}
  \caption{Benchmarking of methods for partial alignment of a single subunit density map to the whole structure (see Dataset section \ref{sec:dataset}). We performed the alignment with each method $50$ times and recorded the angle difference between the output and the ground truth, and RMSD of paired atoms. We reported here the mean and std, with the best results for each measurement highlighted in bold (violin plots and examples are shown in Supplementary Figure S2).}
  \label{table:empot}
  \centering
  
\resizebox{1\textwidth}{!}{
  \begin{tabular}{l|ccccc}
    \toprule
    \cmidrule(r){1-5}
    Metric  & EMPOT (ours)  & AlignOT &  BOTalign & EMAlign & ChimeraX  \\
    \midrule
    Angle difference (${}^\circ$)  & $\mathbf{4.22 \pm 5.74}$   & $20.15 \pm 17.63$  & $10.16 \pm 1.16$  & $42.61 \pm 65.43$ & $99.15 \pm 49.60$\\
    \midrule
    RMSD & $\mathbf{5.20 \pm 7.54}$& $255.41 \pm 6.99$  & $252.00 \pm 4.99$  & $128.32 \pm 130.88$ & $311.38 \pm 150.42$\\
    \bottomrule
  \end{tabular}}
\end{table}

\subsection{Application for atomic model building}

\begin{figure}
  \centering
  \includegraphics[width=1\linewidth,trim={0cm 4.5cm 0cm 0cm} ,clip]
  {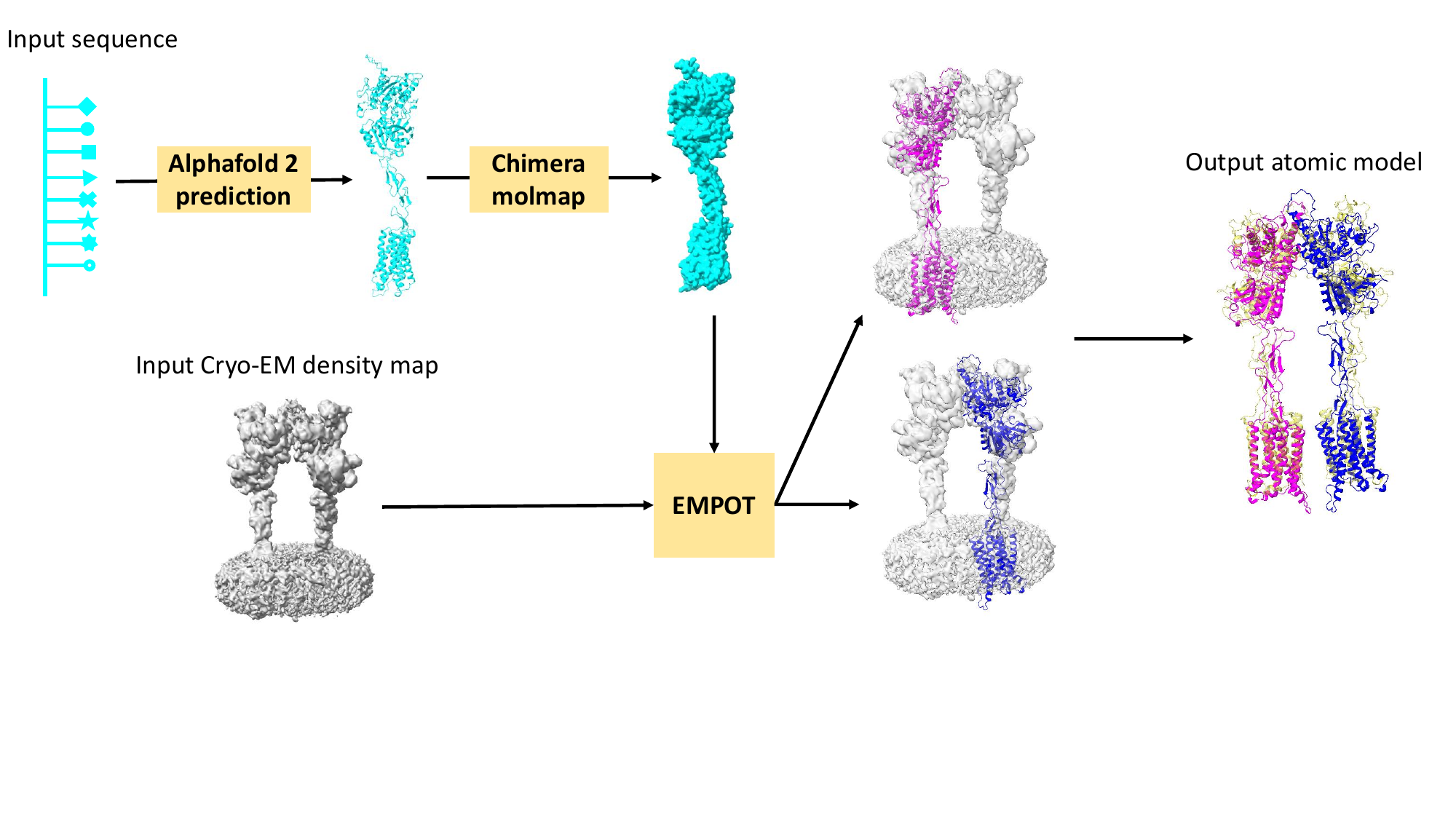}
  \caption{Illustration of the atomic model building procedure using EMPOT for EMD-0346 (PDB ID: 6N52). Structures of the two identical subunits were first obtained using AlphaFold2, and subsequently converted into density maps with Chimera. Partial alignment of the maps was performed with EMPOT, with the output atomic model obtained by concatenating the two fitted subunits (blue and pink, with the ground truth structure from PDB shown in yellow).}
  \label{fig:atom}
\end{figure}

We next study another potential application of our method for atomic model building, by simply extending the framework of our previous experiment. In this case, we started from the sequence associated with the subunit and generated a structure from AlphaFold2. We then used the \texttt{molmap} function in UCSF ChimeraX \cite{chimeraX} to transform this generated model into a density map at a resolution matching the target map (4\AA). The two maps were then aligned using EMPOT. Specifically, we ran the point cloud representation with random initialization so that one of the homologous subunits could fit into either side of the target map (as illustrated in the second right column of Figure \ref{fig:atom}), and we repeated the procedure to get a global fitted atomic model with the two subunits. Figure \ref{fig:atom} summarizes the procedure, with the output atomic model showing that the built model matches quite well with the deposited structure, despite the presence of the micelle and some discrepancies between the AlphaFold structure and the ground truth. We also note that to ensure a successful alignment we increased the point cloud size (from 500 to 2000), and increased the runtime of the procedure to 13 minutes on an AWS EC2 G5 instance with 16 vCPUs.

\begin{table}
  \caption{Comparison of the built models against the deposited reference PDB structure for EMPOT, phenix.dock\_in\_map, gmfit, and DEMO-EM on the EM density map, EMD-0346. TM-score provides a measurement of similarity between atomic models comprised between 0 and 1, with 1 indicating perfect match. Best results for each metric are highlighted in bold.}
  \label{table:atom}
  \centering
  \begin{tabular}{l|cccc}
    \toprule
    \cmidrule(r){1-5}
    Metric  & EMPOT (ours)  & phenix.dock\_in\_map & gmfit & DEMO-EM  \\
    \midrule
    TM-score & \textbf{0.722}  &  0.701 & 0.670  & 0.551  \\
    \bottomrule
  \end{tabular}
\end{table}

To quantitatively evaluate and benchmark the accuracy of the protein complex model built from our method, we adopted the TM-score metric calculated by MM-align \cite{mmalign}, that measures closeness between the built atomic model and the corresponding ground truth PDB structure.
We benchmarked EMPOT against two rigid fitting tools phenix.dock\_in\_map \cite{phenix} and gmfit \cite{gmfit}, and one flexible fitting tool called DEMO-EM \cite{demoem} using the TM-score \cite{mmalign} for evaluating their performance. The results, shown in Table \ref{table:atom}, indicate that EMPOT achieved the best TM-score of 0.722 (1 giving perfect match) while significantly outperforming the other methods tested, suggesting EMPOT as a potential valuable alternative for rigid body fitting.  


\section{Discussion}

In this paper, we present EMPOT, a new method for partial alignment of cryo-EM density maps that relies on minimizing the Gromov-Wasserstein divergence between sampled point clouds. Our experiments suggest that EMPOT is scalable to the typical 
size of density maps, and can be more specifically used for partial alignment, as this case proves to be challenging for other methods. In this regard, our method also extends the framework of classical OT based methods, that have recently been introduced for various problems in Cryo-EM (\cite{ecoffet2020morphot, ecoffet2022application,  zelesko2020earthmover, rao2020wasserstein,riahi2022alignot,singer2023alignment}). In the context of Cryo-EM, solving partial optimal transport allows EMPOT to be especially relevant with the increasing ability of cryo-EM to solve large complex macromolecules with multiple subunits. In addition, we also demonstrate that our method can be used for rigid body fitting, and thus be potentially integrated to pipelines for building atomic models, as a preliminary step to flexible fitting and refinment methods. 

While our results seem to indicate that the present approach is both simple and appropriate to solve challenging problems inherent with partial alignment of density maps, it would be interesting to further test its ability to handle a variety of cases, with multiple heterogeneous subunits, as well as to evaluate and optimize its computational cost. Investigating the choice of a specific vector quantization method for point cloud generation, or of advanced algorithms to sequentially or jointly register multiple subunits  can improve the accuracy and efficiency of our method. We are currently pursuing these directions.

\clearpage 
\bibliographystyle{unsrt}
\bibliography{bibliography}



\end{document}